# Design of Minimum Correlated, Maximal Clique Sets of One-Dimensional Uni-polar (Optical) Orthogonal Codes

R. C. S. Chauhan[1], MIEEE, Y. N. Singh[2], SMIEEE, R. Asthana[3], SMIEEE

*Abstract:* — **This paper proposes an algorithm to search a family of multiple sets of minimum correlated one dimensional uni-polar (optical) orthogonal codes (1-DUOC) or optical orthogonal codes (OOC) with fixed as well as variable code parameters. The cardinality of each set is equal to upper bound. The codes within a set can be searched for general values of code length *'n'*, code weight *'w'*, auto-correlation constraint less than or equal to $\lambda_a$ , and cross-correlation constraint less than or equal to $\lambda_c$ , such that $n >> w > (\lambda_a , \lambda_c)$ . Each set forms a maximal clique of the codes within given range of correlation properties $(\lambda_a , \lambda_c)$ . These one-dimensionaluni-polar orthogonal codes can find their application as signature sequences for spectral spreading purpose in incoherent optical code division multiple access (CDMA)systems.**

## 1. INTRODUCTION

TO implement a CDMA system, one need a set of CDMA codes having a desired properties. Usually, there exist more than one set of such sets for given parameters. We desire to find the algorithm to identify all the sets of minimum correlated orthogonal codes [1]-[4]. Similarly, in Optical CDMA multiple sets of minimum correlated one-dimensional uni-polar (optical) orthogonal codes with fixed or variable code parameters are required to increase the channel capacity [5]-[6] and inherent security. The code parameters for one dimensionaluni-polarorthogonalcodesarecodelength *'n'*, code weight *'w'*, auto-correlation constraint $\lambda_a$and cross- correlation constraint $\lambda_c$ such that $n>>w>(\lambda_a,\lambda_c)$.

Various one-dimensional optical orthogonal code design schemes for constant weight have been proposed in literature [7]-[13]. These schemes can design single set of optical orthogonal codes corresponding to specific values of code parameters $(n, w, \lambda_a ,\lambda_c )$ . The sets of 1-DUOC with variable or multi-weight parameter have larger cardinality than that of the set with constant code weight parameter [14]. The set of codes with low code weights provide poor BER performance, then the set of codes with large code-weights are desirable. The set of codes having subsets with different code weight parameterscanprovidemultipleQoS(qualityofservice)as per the need [14]-[20]. The sets of 1-DUOC or OOC with variable or multi code-length parameter can be used for multi- rate systems employing OOC [5], [21]-[25]. The 1-DUOC with multi-length and multi-weight provide the multi-class set of 1-DUOC with larger cardinality and inherent security [5], [26]-[27] for use in multi-rate systems. The general values or unspecified parameters of the codes increase the inherent security of the system by decreasing the probability of generating same set of signature sequences (pattern) or orthogonal codes [27], unless code parameters are known. It can be said that the sets of 1-DUOC or OOC with general and variable code parameters are needed for systems incorporating OOC for better performance [5],[6].

We have designed the single family of minimum correlated multiple sets for fixed code parameters through proposed maximal clique search method. Secondly two or more such families can be found for various length and weight parameters. Finally one set from each family is searched such that it has minimum correlation with all others. These finally searched minimum correlated maximal clique sets of orthogonal codes with multi-length and multi-weight parameters even with equal or unequal values of auto- correlation constraint and cross-correlation constraint can be put in other family. The auto-correlation constraint for the set of codes designed here is never greater than two. The cross-correlation constraint for set of codes is always equal to one but this may exceeds to two for multiple sets of codes with fixed or variable code parameters representing tradeoff between larger cardinality and better BER performances. Each set has maximum number of codes which is given by upper bound of the set [28]-[33] such that the codes within every set form a maximal clique. In graph theory, a clique is a sub-graph such that each pair of nodes in the sub-graph is connected or adjacent. We can represent all codes as nodes and a link exist between two nodes if cross-correlation is less than or equal to $\lambda_c$ . A sub-set of codes where each possible pair of codes has a link between them is the clique set [47].

In the next section, the basic properties and characteristics of one dimensional uni-polar (optical) orthogonal codes and their multiple maximal clique sets are discussed in brief which have been proposed already in [5],[6],[28]-[38]. In section III, the difference of position (DoP) representation and extended DoP (EDoP) matrices of the 1-DUOC are explored with the help of references [6],[39]-[40]. The section IV describes the method of finding maximum non-zero shift auto-correlation and cross-correlation values of 1-DUOC through conventional method as well as using EDoP matrices of uni-polar codes. In section V, an algorithm to design minimum correlated, maximal clique sets of 1-DUOC with constant as well as variable code parameters, have been proposed withthe tabulated results and calculation of computational complexity of the algorithm. The last section compares the results with already proposed schemes and algorithms for constructing the sets of 1-DUOC or optical orthogonal codes (OOC). In the end, future scope and applications of the designed codes have beendiscussed.

## 2. ONEDIMENSIONALUNI-POLARORTHOGONALCODESAND THEIR MULTIPLESETS

Let two uni-polar code words $X$ and $Y$ belong to a code set with code parameters $(n,w,\lambda_a,\lambda_c)$.

$X=(x_0,x_1,...,x_{n-1}), Y=(y_0,y_1,...,y_{n-1}); \forall x_t, \forall y_t \in (0,1) \forall t.$

*Definition 2.1: [28]*

The maximum of non-zero shift auto-correlation of uni-polar or binary code X is given as $\lambda_{ax}$ [28].

$\lambda_{ax} \geq \sum_{t=0}^{n-1} x_t x_{t \oplus m} \; for \; 0<m\leq n-1.$

$t \oplus m$ implies $(t+m) \bmod (n)$.

*Example 2.1(a):*

Let the code X with length '$n$'=13 and code weight '$w$'=4, be [0 1 0 1 0 0 1 0 0 0 1 0 0].

For 0 <$m$ ≤ 12, the left circular

shifted binary sequences ($X_1$,$X_2$,…,$X_{12}$) of the code X, are as follows.
X = [0 1 0 1 0 0 1 0 0 0 1 00],
$X_1$= [1 0 1 0 0 1 0 0 0 1 0 00],
$X_2$= [0 1 0 0 1 0 0 0 1 0 0 0 1], … ,
$X_{12}$= [0 0 1 0 1 0 0 1 0 0 0 1 0].
The overlapping of weighted bits or non zero shift auto- correlation of code X with its circular shifted binary sequences($X_1$,$X_2$,…$X_{12}$) are (0,1,1,2,1,1,1,1,1,1,1,0). The maximumof all such values is termed as maximum non-zero shift auto- correlation $\lambda_{ax}$ of the code. It will be 2 in this case.

*Definition 2.2: [28]*

If $X_P$ is weighted positions representation (WPR) [36] of uni- polar orthogonal code X of length '*n*' and weight '*w*', the maximum non-zero shift auto-correlation $\lambda_{ax}$ of the code is given as follows.

$\lambda_{ax} \geq (a + X_P) \cap (b + X_P)$, $(a \neq b)$, $0 \leq (a,b) \leq n - 1$. $X_P$ contains '*w*' integer values showing weighted positions or positions of bit 1's of the code X. Here
$a + X_P = \{(a + x_P) \bmod n : x_P \in X_P\}$.

*Example 2.2(a):*

Let the uni-polar code X = [1 0 1 0 0 1 0 0 0 1 0 0 0] with code-length '*n*'=13, and the code-weight '*w*'=4, has its weighted positions representation $X_P$ = (0,2,5,9). The circular shifted sequences of the code X, or (a+ $X_P$) or (b+ $X_P$) for 0 ≤(*a*,*b*) ≤12, are given as following. [(0,2,5,9),(1,3,6,10), (2,4,7,11), (3,5,8,12), [(0,4,6,9), (1,5,7,10), (2,6,8,11), (3,7,9,12), (0,4,8,10), (1,5,9,11), (2,6,10,12), (0,3,7,11), (1,4,8,12)].
The intersection of these circular shifted weighted position sequences (a+$X_P$) with (b+$X_P$) is not greater than 2. Hence the maximum non-zero shift auto-correlation of the code X is equal to 2.

*Definition 2.3:[6]*

If $X_P$ is weighted positions representation [36] of uni-polar orthogonal code X of length '*n*' and weight '*w*', the maximum non-zero shift auto-correlation $\lambda_{ax}$ of the code is also given as follows.

$\lambda_{ax} \geq (X_P) \cap (a + X_P)$, $(0 < a \leq n - 1)$

*Example 2.3(a):*

Let us take same code X as in examples 2.1(a) and 2.2(a). The intersection of WPR of code X, $X_P$ = (0,2,5,9) with circular shifted sequences of X or (a+$X_P$) is not greater than 2. Hence the maximum non-zero shift auto-correlation X is equal to 2.

*Definition 2.4: [28],[38]*

Auto-correlation constraint $\lambda_a$ for the set of 1-DUOC is always greater than or equal to maximum non-zero shift auto- correlation $\lambda_{ax}$ of every code within the set. $\lambda_a \geq \lambda_{ax}$ .

*Example 2.4(a):*

Let the set of one dimensional uni-polar orthogonal codes is (X,Y,Z,A,B). The maximum non-zero shiftauto-correlation $\lambda_{ax}$ of the codes X,Y,Z,A,B are 1,2,1,2,2 respectively.The auto-correlation constraint $\lambda_a$for the set is maximum of (1,2,1,2,2) i.e., $\lambda_a$

= 2.

*Lemma 2.5: [28],[38]*

For code X, the maximum non-zero shift auto-correlation $\lambda_{ax}$satisfy the following relation, $1 \leq \lambda_{ax} \leq w - 1$, for 1- DUOC with code parameters ($n$, $w \geq 2$).

Proof: In the uni-polar code with $w \geq 2$ , at least oneweighted bit will always overlap with one of the $(n - 1)$ non-zero circular shifted versions. No uni-polar code with its every non- zero circular shifted version results in ' $w$ ' overlapped weighted bits. Because ' $w$ ' overlapping weighted bits occurs only with the codes un-shifted or zero (mod ($n$)) circular shifted versions. Then the maximum overlapping of code with its non-zero circular shifted versions is less than $w$ i.e. less than equal to $(w - 1)$. Hence for the code parameters ($n$, $w \geq 2$) the values of maximum non-zero shift auto- correlation of the codes lies in the range 1 to $(w - 1)$.

*Definition 2.6: [28],[38]*

The maximum cross-correlation of a uni-polar code X with another code Y and all the ($n$ −1) circular shifted versions of code Y is defined as cross-correlation $\lambda_{cxy}$ for the pair of codes X and Y [28],[38],andsatisfies

$\lambda_{cxy} \geq \sum_{t=0}^{n-1} x_t y_{t \oplus m}$ $or$ $\sum_{t=0}^{n-1} y_t x_{t \oplus m}$ for $0 \leq m \leq n - 1$

*Example 2.6(a):*

Let the code length *'n'*=13, code weight '*w*'=4, the uni-polar codeX=[0101001000100]andcodeY=[1101000 001000].Themaximumnon-zeroshiftauto-correlationof both X and Y is 2. The overlapping of weighted bits of code Y with X and all 12 circular shifted versions of code X i.e. ($X_1$,$X_2$,…,$X_{12}$) (as given in example 2.1(a)) are (2,0,1,2,1,0,2,1,2,0,2,1,2). The maximum of these cross correlation values is 2 which is the cross-correlation for the pair of codes X and Y, i.e. $\lambda_{cxy}$= 2 .

*Definition 2.7: [28]*

If $X_P$ and $Y_P$ are weighted positions representation (WPR) [36] ofuni-polarorthogonalcodeXandYrespectivelywithcode- code-length 'n' and weight 'w', the cross-correlation of the pair of code X and Y is given as follows $\lambda_{cxy}$

$\lambda_{cxy} \geq (a + X_P) \cap (b + Y_P)$, $0 \leq (a,b) \leq n - 1$.

*Example 2.7(a):*

Let the code length *'n'*=13, code weight '*w*'=4, the uni-polar code X= [1 0 1 0 0 1 0 0 0 1 0 0 0] and code Y= [1 1 0 1 0 0 0 0 0 1 0 0 0] with its weighted positions representation $X_P$ = (0,2,5,9) and $Y_P$ = (0,1,3,9) respectively. The circular shifted sequences of the code X, or ($a+X_P$) with its weighted positions are given as following.

[(0,2,5,9), (1,3,6,10), (2,4,7,11), (3,5,8,12)], (0,4,6,9), (1,5,7,10),(2,6,8,11), (3,7,9,12), (0,4,8,10), (1,5,9,11), (2,6,10,12), (0,3,7,11), (1,4,8,12)]

The circular shifted sequences of the code Y, or ($b+Y_p$) with its weighted positions are given asfollowing.

[(0,1,3,9), (1,2,4,10), (2,3,5,11), (3,4,6,12), (0,4,5,7),(1,5,6,8), (2,6,7,9), (3,7,8,10), (4,8,9,11), (5,9,10,12), (0,6,10,11), (1,7,11,12), (0,2,8,12)]

The intersection of these circular shifted sequences ($a+X_P$) and ($b+Y_P$) with its weighted positions is not greater than 2. Hence the cross-correlation $\lambda_{cxy}$

*Definition 2.8: [6]*

If uni-polar code X and Y of length '$n$' and weight '$w$' are represented with its '$w$' weighted positions, the cross- correlation $\lambda_{cxy}$ of the code is also given as follows

$$\lambda_{cxy} \geq (X_P) \cap (a + Y_P), (0 \leq a \leq n-1)$$

Alternatively

$$\lambda_{cxy} \geq (Y_P) \cap (a + X_P), (0 \leq a \leq n-1)$$

*Example 2.8(a):*

Let the code length be '$n$'=13, code weight '$w$'=4, the uni- polar code X= [1 0 1 0 0 1 0 0 0 1 0 0 0] and code Y= [1 1 0 1 000001000]withtheirweightedpositionsrepresentation $X_P$=(0,2,5,9)and$Y_P$=(0,1,3,9)respectively.Thecircular shifted sequences of the code Y, or (a+$Y_P$) are given as in example 2.7(a), (b). The intersection of code $X_P$ and the circular shifted sequences (a+$Y_P$) with its weighted positions is not greater than 2. Hence the cross-correlation $\lambda_{cxy}$ of the code X and code Y is equal to 2.

*Definition 2.9: [28],[38]*

The cross-correlation constraint $\lambda_c$ for the set of 1-DUOCs is always greater than or equal to cross-correlation $\lambda_{cxy}$ of any pair of codes within the set . $\lambda_c \geq \lambda_{cxy}; \forall x, y$.

*Example 2.9(a):*

Let the set of 1-DUOCs be (X,Y,Z,A,B). The pairs of codes within set are (XY,XZ,XA,XB,YZ,YA,YB,ZA,ZB,AB). Let the cross-correlation values for these pairs of codes are (2,1,2,2,1,1,2,1,1,2) respectively. The cross-correlation constraint $\lambda_c$ for the set is maximum of (2,1,2,2,1,1,2,1,1,2), i.e. $\lambda_c$ = 2 for the set of codes (X,Y,Z,A,B).

*Lemma 2.10: [28],[38]*

For the pair of 1-DUOC with code- parameters ($n$, $w \geq 2$) , X and Y, the cross-correlation $\lambda_{cxy}$ satisfies the following relation, $1 \leq \lambda_{cxy} \leq w-1$. Proof: In a pair of uni-polar codes with code parameters ($n$, $w \geq 2$) , at least one weighted bit of one uni- polar code will always overlapped with other code or one of the $(n-1)$ non-zero circular shifted versions of othercode.

Further no uni-polar code will results in ' $w$ ' overlapping of weighted bits with other code or non-zero circular shifted versions of other code. Because ' $w$ ' overlapping of weighted bits occurs only with its own un-shifted or zero (mod ($n$)) circular shifted version. Thus the maximum overlapping of code with other code or non-zero circular shifted versions of other code may result in less than $w$ or less than equal to $(w-1)$ overlapping. Hence, for the code parameters ($n$, $w \geq 2$) , the cross-correlation of the pair of codes lies between 1 to $(w-1)$. The one-dimensional uni-polar orthogonal codes with $\lambda_{cxy}$= 1 are perfectuni-polar orthogonal codes, while the codes with $1 < \lambda_{cxy}$ are quasi orthogonal.

*Theorem 2.11:*

The orthogonality and cardinality of the maximal set of one- dimensional uni-polar orthogonal codes are inversely proportional to each other.

Proof:

The pair of uni-polar codes with $\lambda_c$ = 1, is termed as maximum orthogonal 1-DUOC pair. While the pair of uni- polar codes with $\lambda_c = w - 1$, is termed as minimum orthogonal pair of 1-DUOC.

For $\lambda_a = \lambda_c = \lambda$ where $1 \leq \lambda \leq w - 1$, the maximum number of one dimensional uni-polar orthogonal codes Z, within a set, is given by following Johnson bound[28]-[31],

$$Z(n, w, \lambda) \leq \left\lfloor \frac{1}{w} \left\lfloor \frac{n-1}{w-1} \dots\dots\dots\dots \left\lfloor \frac{n-\lambda}{w-\lambda} \right\rfloor \right\rfloor \right\rfloor = J_A(n, w, \lambda)$$

Here $\lfloor a \rfloor$ representslargestintegerlessthanequaltoa. For $\lambda = w - 1$

$$Z(n, w, w-1) \leq \left[\frac{1}{n} {}^{n}C_w\right] = \left\lfloor \frac{1}{w} \left\lfloor \frac{n-1}{w-1} \dots\dots\dots\dots \left\lfloor \frac{n-(w-1)}{1} \right\rfloor \right\rfloor \right\rfloor$$

which represent maximum number of 1-DUOCs within one set with minimumorthogonality.

$$For\ \lambda = 1, Z(n, w, 1) \leq \left\lfloor \frac{1}{w} \left\lfloor \frac{n-1}{w-1} \right\rfloor \right\rfloor$$

which represents to minimum number of uni-polar orthogonal codes in one set with maximum orthogonality.

For ($\lambda = p$), ($1 < p < w - 1$) , the cardinalityof maximal set is

$$Z(n, w, p) \leq \left\lfloor \frac{1}{w} \left\lfloor \frac{n-1}{w-1} \dots\dots\dots\dots \left\lfloor \frac{n-p}{w-p} \right\rfloor \right\rfloor \right\rfloor$$

which is less than the cardinality of maximal set for , $\lambda = (p + 1)$

$$Z(n, w, p+1) \leq \left\lfloor \frac{1}{w} \left\lfloor \frac{n-1}{w-1} \dots\dots\dots\dots \frac{n-p}{w-p} \left\lfloor \frac{n-(p+1)}{w-(p+1)} \right\rfloor \right\rfloor \right\rfloor$$

While the orthogonality for the set with ($\lambda = p$) is greater than for the set with ($\lambda = p + 1$) . It proves that orthogonality and cardinality of maximal set are inversely related to each other.

*Lemma 2.12:*

The maximal set of 1-DUOCs with parameters ($n$, $w$, $\lambda_a$ , $\lambda_c$ ) forms a maximal clique of codes.

Proof:

All the codes in a set are such that every pair of codes is having correlation properties within given range. If the codes are assumed to be nodes, then each node is connected with all others with the given properties. This shows that all the codes within set form a clique. If the cardinality of the set is maximum or equal to upper bound; it means that the formed clique of codes is maximal. A code is chosen and we can keep on adding another code to extend the set so that extended set is a clique. Once it is no more possible to extend the set further, we have achieved a maximal clique.

*Theorem 2.13:*

For the code parameters ($n$, $w$, $\lambda_a$ , $\lambda_c$ ) , the cardinality of maximal clique set and number of maximal clique sets are inversely proportional to each other.
Proof:
As per Theorem 2.1, a single set of the 1-DUOC is possible with minimum orthogonality or ($\lambda = w - 1$) and maximum cardinality. Moreover, for ($\lambda = p$) , ($1 \leq p < w - 1$) , the cardinality of the maximal set is less than for ($\lambda = p + 1$) . Then more codes are available for forming more sets for ($\lambda = p$) than for ($\lambda = p + 1$) . It proves that cardinality of maximal set and numbers of maximal sets are inversely proportional to each other.

*Lemma 2.14:*

The minimum cross-correlation among the multiple maximal clique sets for the code parameters ($n$, $w$, $\lambda_a$ , $\lambda_c$ ) is equal to ($\lambda_c + 1$) .

Proof:

Forthecodeparameters($n$,$w$,$\lambda_a$,$\lambda_c$),themaximalclique set contains the codes with auto-correlation constraint less than or equal to $\lambda_a$and cross-correlation constraint less than or equal to $\lambda_c$ . The cross correlation between two independentmaximalcliquesetsisequaltomaximumcross- correlation for the pair of codes. One code is taken from one set and other one from the second. This maximum cross-correlation cannot be less than or equal to $\lambda$because boththe sets are maximal. It will always be greater than $\lambda_c$ . Hence the minimum value of the cross-correlation among the multiple independent maximal clique sets is equal to ($\lambda_c + 1$) . This also implies that no code shall be common between two maximal clique sets. If such a code exist, cross correlation between codes taken from two sets will be less than equalto
$\lambda_c$ and thus sets are not maximal clique sets.

To design multiple maximal sets of codes with general values of code parameters ($n$, $w$, $\lambda_a$ , $\lambda_c$ ) , a new method is proposed with difference of position representation (DoPR) and calculation of correlation values. Before discussing the method, the characteristics of difference of positions representation for the one dimensional uni-polar orthogonal codes is discussed in the nextsection.

## 3. DIFFERENCE OF POSITIONS REPRESENTATION (DOPR)

Conventionally optical orthogonal codes are represented with their weighted positions [6],[7]-[14], which is not a unique representation of the code because weighted positions always change with circular shift of the code. One- dimensional uni-polar orthogonal codes are assumed to be the same with every circular shift of the code [14] for asynchronous use of the code in the multiple access systems. The difference of positions representation ( DoPR) of the code remains same even with circular shift of the code. The DoPR is taken from difference families of optical orthogonal codes discussed in [6],[39]-[40].

*Lemma 3.1:*
The *'w'* differences of consecutive weighted positions of one-dimensional uni-polar orthogonal code remain unchanged for every circular shift of the uni-polar code [14].
Proof:

Theuni-polarcodeXwithcodelength*‘n’*andweight*‘w’* has‘*w’*weightedpositions.ThebinarycodeXcanbeputon the periphery of the circle in serial order so that last and first bits of the code are adjacent. Now on every circular shift of the binary code around the circle, the difference of secondandfirst weighted position remains same. Similarly for every circular shift of the code, the difference between $(j+1)^{th}$ and $j^{th}$ weighted positions also remains same. Finally, it can be observed that all the *‘w’* differences of consecutive weighted positions of the code remains unchanged on every circular shift of the code. Here $(j) \leq w$ and difference iscalculated under modulo $n$ arithmetic.

The lemma 3.1 gives the idea for unique representation of the code having *‘w’* differences of consecutive weighted positions of the code. These *‘w’* difference of consecutive weighted positions of the code is termed as difference of position representation (DoPR) of the code. There are *‘w’* or less than *‘w’* circular shifted DoPR of the code. One of these circular shifted DoPR can be standardized to represent the codeuniquely.

*Example 3.1(a):*

Let us take the code X =[0 1 0 1 0 0 1 0 0 0 1 0 0] with its WPR, $X_P$ = (1,3,6,10). The differences of consecutive weighted positions of the code are (2,3,4,4) under modulo n = 13 arithmatic. For every circular shifted version of code X, ($X_1$,$X_2$,…,$X_{12}$), the differences of consecutive weighted positions of these shifted version remain un-changed and these will be (2,3,4,4) or (3,4,4,2) or (4,4,2,3) or (4,2,3,4).The DoPR of the code X is (2,3,4,4) and the circular shifted DoPR of the code are (3,4,4,2), (4,4,2,3),(4,2,3,4).

*Lemma 3.2:*

The difference of any two weighted positions of the uni- polar code always lies from oneto ($n$-1).  *Theorem 3.3:[6]*

The sum of all the *‘w’* differences of consecutive weighted positions or the elements of DoPR of the uni-polar code is always equal to code length *‘n’* .

Proof:

For the WPR of uni-polar code X, $X_P = (x_{p1}, x_{pi}, x_{pj} \ldots x_{pw})$ .

First difference $d_{x1}$ of positions $(x_{p1}, x_{pi}) = (x_{pi} - x_{p1})$

Second difference of positions $(x_{pi}, x_{pj})$, $d_{x2} = (x_{pj} - x_{pi})$ $(w\text{-}1)^{th}$ difference of positions $(x_{p(w-1)}, x_{pw})$, $d_{x(w-1)} = (x_{pw} - x_{p(w-1)})$ $(w)^{th}$ difference of positions $(x_{pw}, x_{p1})$, $d_{xw} = (n + x_{p1} - x_{pw})$ the $(w)^{th}$ difference $d_{xw}$ is calculated under modulo *‘n’* arithmetic because $(x_{p1} < x_{pw})$

The sum of all *‘w’* differences=$(d_{x1} + d_{x2} + \ldots + d_{xw}) = \Big((x_{pi} - x_{p1}) + (x_{pj} - x_{pi}) + \cdots .. + (x_{pw} - x_{p(w-1)}) + (n + x_{p1} - x_{pw})\Big) = n$

*3.4 Formation of Standard DoPR of the Code*

The one-dimensional uni-polar orthogonal code has a proper representation as DoPR containing *‘w’* differences of consecutive positions (DoPs). The uni-polar code can be represented by any one of the *‘w’* circular shifted DoPR. One of these ‘w’ circular shifted DoPR can be fixed as standard DoPR following the procedure givenbelow.

*Step 1.* Out of the ‘w’ circular shifted DoPR, the DoPR with last element greater than other

(w-1) DoPs, is selected as standard DoPR of the code.

*Example 3.4(a):*

Let the uni-polar code with code length n=31, weight w=5, be (2,5,13,4,7) in DoPR. The circular shifted DoPRs are (5,13,4,7,2), (13,4,7,2,5), (4,7,2,5,13), and (7,2,5,13,4). The standard DoPR of the code is (4,7,2,5,13) which has highest element as last one.

*Step 2.* If after the step '1', the code has more than one DoPR with highest last element but equal to some DoPs of that DoPR, the DoPR with smallest value of first DoP element, is selected as standard DoPR of the code.

*Example 3.4(b):*

Let the uni-polar code with code length n=31, weight w=5, be (6,6,7,5,7) in DoPR. The other circular shifted DoPRs of the code are (6,7,5,7,6), (7,5,7,6,6), (5,7,6,6,7),(7,6,6,7,5). The

DoPRs selected after step 1 for standard DoPR are (6,6,7,6,7) and (5,7,6,6,7). The standard DoPR of the code is (5,7,6,6,7) with smaller first element.

*Step 3.* If in the step '2' we get more than one DoPR with highest last and smallest first DoPs, the DoPR with smaller value of second DoP, is selected as standard DOPR.

*Example 3.4(c):*

Let the uni-polar code with code length n=31, weight w=5, is (6,5,7,6,7) in DoPR. The other circular shifted DoPRs of the code are given as follows (5,7,6,7,6), (7,6,7,6,5), (6,7,6,5,7), (7,6,5,7,6). The DoPRs selected from step 1 for standard DoPR are (6,5,7,6,7) and (6,7,6,5,7). The step 2 could not standardize the code from two DoPR (6,5,7,6,7) and (6,7,6,5,7) of the code because first element of both DoPR is same and equal to 6. The step third results in the standard DoPR of the code as (6,5,7,6,7) with smaller second DoP element out of both circular shiftedDoPRs.

*Step 4.* The process may continue till unique and standard DoPR of the code is found, by comparing third, fourth and so on elements in same fashion.

*Lemma 3.5:*

In the standard DoPR of the unipolar code of length '*n*' and weight '*w*', the range of first $\left[\frac{w-1}{2}\right]$DoP elements lies from 1 to$\left[\frac{n-w+1}{2}\right]$while the range of next$\left[\frac{w-1}{2}\right]$DoP elements lies from 1 to$\left[\frac{n-w+2}{2}\right]$

Proof: Let the standard DoPR of the uni-polar code is $(d_{x1},d_{x2},...,d_{xw})$. The minimum values of $(d_{x1},d_{x2},...,d_{x(w-1)})$ are equal to 1 as per lemma 3.2. The first DoP element $(d_{x1})$ takes its maximum value when$(d_{x2}=d_{x3}=...=d_{x(w-1)})=1$and$(d_{xw}>d_{x1})$ or $d_{xw}=d_{x1}+1)$ for standard DoPR. As per Theorem 3.3,

$$(d_{x1}+d_{x2}+...+d_{x(w-1)}+d_{xw})=n$$

$$(d_{x1}+1+...+1+(d_{x1}+1))=n$$

$$(d_{x1}+d_{x1})=n-(w-1)$$

$$d_{x1}=\lfloor(n-w+1)/2\rfloor.$$

Similarly$(d_{x2})$oroneoffirst$\lfloor(w-1)/2\rfloor$DoPelements$(d_{xi})$,$(1\leq i\leq\lfloor(w-1)/2\rfloor)$,takesitsmaximumvaluewhen other DoPelements except $(d_{xw})$ equal to oneand $(d_{xw}>d_{xi})$ or $(d_{xw}=d_{xi}+1)$ so that

$d_{xi}=\lfloor(n-w+1)/2\rfloor$forstandardDoPROneofthenextremaining$\lceil(w-1)/2\rceil$DoPelements except

last DoP element ($d_{xj}$), ($\lfloor (w-1)/2 \rfloor < j \leq (w-1)$), takesmaximumvalue when other DoP elements except ($d_{xw}$) equal to one and ($d_{xw} \geq d_{xj}$) or ($d_{xw} = d_{xj}$) for standard DoPR.

Asper Theorem 3.3,

$$(d_{x1}+d_{x2}+...+d_{x(w-1)}+ d_{xw}) = n$$
$$(d_{xj}+(w-2)+d_{xj})=n$$
$$2d_{xj}=n-(w-2)$$
$$d_{xj}=\lfloor (n-w+2)/2 \rfloor.$$

If one of the first $\lfloor (w-1)/2 \rfloor$ DoPelementsequaltolastDoP element and no element of second half $\lceil (w-1)/2 \rceil$ DoP elements equal to last DoP element, the code can be standardized by taking one of its circular shifted versions such thatfirst $\lfloor (w-1)/2 \rfloor$ DoPelementshavenoDoPelement equal to last DoP element.

*Lemma 3.6:*

In the standard DoPR of the uni-polar code of length '$n$' and weight'$w$',lastDoPelementisinrangefrom $\lceil n/w \rceil$ to $(n - w + 1)$ .

Proof: Suppose the standard DoPR of the uni-polar code is($d_{x1}$, $d_{x2}$ ,..., $d_{xw}$) . The last DoP element ($d_{xw}$) takes its maximum value when all other DoP elements are minimum or equal to one. Then maximum of ($d_{xw}$) is equal to($n-w+1$)as per theorem 3.3. This ($d_{xw}$) takes its minimum value when all other DoP elements are such that their DoP values are just less than or equal to last DoP element.MathematicallysomeofotherDoPelementsare equalto $\lfloor n/w \rfloor$,someare $\lceil n/w \rceil$.Theminimumvalueof lastDoPelement($d_{xw}$)willbe $\lceil n/w \rceil$

The maximum non-zero shift auto-correlation and cross- correlation values of the codes can be calculated using the DoPR or standard DoPR. This calculation is easier than the conventional calculation of auto and cross-correlation values ofthecodesasgivenindefinitions2.1,2.2,2.3,2.6,2.7&2.8. For the calculation of correlation values, the DoPR is converted into extended DoP matrix of the code. The extended DoP matrix ($w \times (w-1)$) of the code contains not only differences of consecutive weighted positions but also the differences of any two weighted positions of the code.

Extended DoP (EDoP) Matrix of the Uni-polar Code.

1. There are 'w' rows and (w-1) columns in extended DoP matrix of thecode.
2. The first row of extended DoP matrix contains differences of first with all other weighted positions of thecode.
3. The $w^{th}$ row of extended DoP matrix contains the differences of $w^{th}$ with all other weighted positions of the code in cyclicorder.

In $j^{th}$ row, the difference of $i^{th}$ element with $(i+1)^{st}$ element can be placed in any column and remaining elements are placed in cyclic order. This mean for same code, we can have$(w-1)^{w}$ EDoP matrices. One of which may be given as follows.

Let us take the code X with DoPR($d_{x1}$,$d_{x2}$,...,$d_{xw}$) with weight '$w$' and code length $n = d_{x1} + d_{x2} + ... + d_{xw}$, theEDoP matrix is formed as following

$$\begin{bmatrix} e_{x01} & e_{x02} & \cdots & e_{x0(w-2)} & e_{x0(w-1)} \\ e_{x11} & e_{x12} & \cdots & e_{x1(w-2)} & e_{x1(w-1)} \\ \cdots & \cdots & \cdots & \cdots & \cdots \\ e_{x(w-2)1} & e_{x(w-2)2} & \cdots & e_{x(w-2)(w-2)} & e_{x(w-2)(w-1)} \\ e_{x(w-1)1} & e_{x(w-1)2} & \cdots & e_{x(w-1)(w-2)} & e_{x(w-1)(w-1)} \end{bmatrix}$$

with

$e_{x01} = d_{x1};\ e_{x11} = d_{x2};\ldots;\ e_{x(w-2)1} = d_{x(w-1)};\ e_{x(w-1)1} = d_{xw};$

$e_{x02} = d_{x1} + d_{x2};\ e_{x12} = d_{x2} + d_{x3};$

$\ldots;$

$e_{x(w-2)2} = d_{x(w-1)} + d_{xw};\ e_{x(w-1)2} = d_{xw} + d_{x1};$

$\ldots;$

$e_{x0(w-2)} = d_{x1} + d_{x2} + \ldots + d_{x(w-2)};\ e_{x1(w-2)} = d_{x2} + d_{x3} + \ldots + d_{x(w-1)};$

$\ldots;$

$e_{x(w-2)(w-2)} = d_{x(w-1)} + d_{xw} + d_{x1} + d_{x2} + \ldots + d_{x(w-4)};$

$e_{x(w-1)(w-2)} = d_{xw} + d_{x1} + d_{x2} + \ldots + d_{x(w-3)};$

$e_{x0(w-1)} = d_{x1} + d_{x2} + \ldots + d_{x(w-1)};e_{x1(w-1)} = d_{x2} + d_{x3} + \ldots + d_{xw};$

$\ldots;$

$e_{x(w-2)(w-1)} = d_{x(w-1)} + d_{xw} + d_{x1} + d_{x2} + \ldots + d_{x(w-3)};$

$e_{x(w-1)(w-1)} = d_{xw} + d_{x1} + d_{x2} + \ldots + d_{x(w-2)}.$

*Example 3.7(a):*

Let the DoPR of the code with weight 'w' equal to 5 is (a,b,c,d,e) and code length '*n'*=a+b+c+d+e. The extended DoP matrix (5x4) is given as

$$\begin{bmatrix} a & a+b & a+b+c & a+b+c+d \\ b & b+c & b+c+d & b+c+d+e \\ c & c+d & c+d+e & c+d+e+a \\ d & d+e & d+e+a & d+e+a+b \\ e & e+a & e+a+b & e+a+b+c \end{bmatrix}.$$

*Lemma 3.8:*

If '*a*' is a DoP element of extended DoP matrix of the code, thentheDoPelement'*n-a*'alsoexistinthesameextended DoP matrix ofthecode.

Proof: if '*a*' is a difference of any twoweightedpositions($x_{pi}$,$x_{pj}$)ofthecodesuchthat$(i,j) \in (0{:}n-1)$

i.e. $a = (x_{pj} - x_{pi})$ whilethe difference between($x_{pj}$, $x_{pi}$ ) in circular order is $(x_{pi} - x_{pj}) = (n - a)$ in modulo 'n' arithmetic. It means that two DoP elements '*a*' and'*n-a*' represents the difference of two same weighted positions.

*Lemma 3.9:*

If first $(w-1)$ consecutive differences of weighted positions EDoP

$$\text{EDoP}\begin{bmatrix} e_{x01} & e_{x02} & \cdots & e_{x0(w-2)} & e_{x0(w-1)} \\ e_{x11} & e_{x12} & \cdots & e_{x1(w-2)} & e_{x1(w-1)} \\ \cdots & \cdots & \cdots & \cdots & \cdots \\ e_{x(w-2)1} & e_{x(w-2)2} & \cdots & e_{x(w-2)(w-2)} & e_{x(w-2)(w-1)} \\ e_{x(w-1)1} & e_{x(w-1)2} & \cdots & e_{x(w-1)(w-2)} & e_{x(w-1)(w-1)} \end{bmatrix}$$

With

$$e_{x01}=d_{x1};\ e_{x11}=d_{x2};\ldots;e_{x(w-2)1}=d_{x(w-1)};$$

$$e_{x(w-1)1}=d_{xw}=(n-(d_{x1}+d_{x2}+\ldots+d_{x(w-1)}));$$

$$e_{x02}=d_{x1}+d_{x2};\ e_{x12}=d_{x2}+d_{x3}\ ;\ldots;$$

$$e_{x(w-2)2}=d_{x(w-1)}+d_{xw}=(n-(d_{x1}+d_{x2}+\ldots+d_{x(w-2)}));$$

$$e_{x(w-1)2}=d_{xw}+d_{x1}=(n-(d_{x2}+d_{x3}+\ldots+d_{x(w-1)}));\ldots;$$

$$e_{x0(w-2)}=d_{x1}+d_{x2}+\ldots+d_{x(w-2)};$$

$$e_{x1(w-2)}=d_{x2}+d_{x3}+\ldots+d_{x(w-1)};$$

$$e_{x(w-2)(w-2)}=d_{x(w-1)}+d_{xw}+d_{x1}+d_{x2}+\ldots+d_{x(w-4)}=(n-(d_{x(w-3)}+d_{x(w-2)}));$$

$$e_{x(w-1)(w-2)}=d_{xw}+d_{x1}+d_{x2}+\ldots+d_{x(w-3)}=(n-(d_{x(w-2)}+d_{x(w-1)}));$$

$$e_{x0(w-1)}=d_{x1}+d_{x2}+\ldots+d_{x(w-1)};e_{x1(w-1)}=d_{x2}+d_{x3}+\ldots+d_{xw}=(n-d_{x1});\ldots;$$

$$e_{x(w-2)(w-1)}=d_{x(w-1)}+d_{xw}+d_{x1}+d_{x2}+\ldots+d_{x(w-3)}=(n-d_{x(w-2)});$$

$$e_{x(w-1)(w-1)}=d_{xw}+d_{x1}+d_{x2}+\ldots+d_{x(w-2)}=(n-d_{x(w-1)}).$$

*Example 3.9(a):*

Let the DoPR of the code with weight 'w' equal to 5 is (a,b,c,d,e) and code length *'n'*=a+b+c+d+e. The extendedDoPmatrix (5x4) is given as

$$\begin{bmatrix} a & a+b & a+b+c & a+b+c+d \\ b & b+c & b+c+d & n-a \\ c & c+d & n-(a+b) & n-b \\ d & n-(a+b+c) & n-(b+c) & n-c \\ n-(a+b+c+d) & n-(b+c+d) & n-(c+d) & n-d \end{bmatrix}$$

*Lemma 3.10:*

If first $u<w$ consecutive differences of weighted positions or DoP element ($d_{x1}$,$d_{x2}$,...,$d_{xu}$)of DoPR($d_{x1}$,$d_{x2}$,...,$d_{x(w-1)}$,$d_{xw}$)ofthecodeare known,the extended DoP matrix for the incomplete code with $u$DoP or DoP element ($d_{x1}$, $d_{x2}$,...,$dx(w-1)$ )of DoPR($d_{x1}$,$d_{x2}$,...,$d_{x(w-1)}$,$d_{xw}$)ofthecodeare known,the extended DoP matrix is given asfollows.

$$\text{EDoP}\begin{bmatrix} e_{x01} & e_{x02} & \cdots & e_{x0(u-1)} & e_{x0u} \\ e_{x11} & e_{x12} & \cdots & e_{x1(u-1)} & e_{x1u} \\ \cdots & \cdots & \cdots & \cdots & \cdots \\ e_{x(u-1)1} & e_{x(u-1)2} & \cdots & e_{x(u-1)(u-1)} & e_{x(u-1)u} \\ e_{x(u)1} & e_{x(u)2} & \cdots & e_{x(u)(u-1)} & e_{x(u)(u)} \end{bmatrix}$$

With

$$\begin{bmatrix} e_{x01} = d_{x1}; \\ e_{x11} = n - d_{x1} \end{bmatrix}$$

$$\begin{bmatrix} e_{x02} = d_{x1} + d_{x2}; e_{x12} = d_{x2} \\ e_{x21} = n - (d_{x1} + d_{x2}); e_{x22} = n - d_{x2} \end{bmatrix}$$

$$\begin{bmatrix} e_{x03} = (d_{x1} + d_{x2} + d_{x3}); e_{x13} = (d_{x2} + d_{x3}); e_{x23} = d_{x3}; \\ e_{x31} = n - (d_{x1} + d_{x2} + d_{x3}); e_{x32} = n - (d_{x2} + d_{x3}); e_{x33} = n - d_{x3} \end{bmatrix}$$

$;\ldots;$

$$\begin{bmatrix} e_{x0(u)} = (d_{x1} + d_{x2} + \ldots + d_{xu}); e_{x1(u)} = (d_{x2} + d_{x3} + \ldots + d_{xu}); \\ e_{x2(u)} = (d_{x3} + d_{x4} + \ldots + d_{xu}); \ldots; e_{x(u-1)(u)} = (d_{xu}); \\ e_{xu1} = n - (d_{x1} + d_{x2} + \ldots + d_{xu}); e_{xu2} = n - (d_{x2} + d_{x3} + \ldots + d_{xu}) \\ ;\ldots; e_{x(u)(u-1)} = n - (d_{x(u-1)} + d_{xu}); e_{x(u)(u)} = n - (d_{xu}); \end{bmatrix}$$

Proof: It is obvious that there is no change in EDoP matrix if EDoP element in same row move to another position. The EDoP matrix of Lemma 3.9 is same as EDoP matrix of lemma
3.10 with cyclic shift within rows. It can be verified by examples 3.9(a) and 3.10(a) with $i^{th}$ row being cyclically shifted left (i-1) times.

The advantage of this kind of EDoP representation is that if last $k$ entries of DoPR are deleted, EDoP can be determined by deleting lower $k$ rows and rightmost k column in EDoP representation of complete code.

*Example 3.10(a):*

Let the DoPR of the code with weight 'w' equal to 5 is
$(a,b,c,d,e)$and code length $n=(a+b+c+d+e)$. The extended DoP matrix for 1, 2, 3, and 4 consecutive DoP elements of incomplete and complete code is given as following matricesrespectively.

$$\begin{bmatrix} a \\ n-a \end{bmatrix},$$

$$\begin{bmatrix} a & a+b \\ n-a & b \\ n-(a+b) & n-b \end{bmatrix},$$

$$\begin{bmatrix} a & a+b & a+b+c \\ n-a & b & b+c \\ n-(a+b) & n-b & c \\ n-(a+b+c) & n-(b+c) & n-c \end{bmatrix},$$

$$\begin{bmatrix} a & a+b & a+b+c & a+b+c+d \\ n-a & b & b+c & b+c+d \\ n-(a+b) & n-b & c & c+d \\ n-(a+b+c) & n-(b+c) & n-c & d \\ n-(a+b+c+d) & n-(b+c+d) & n-(c+d) & n-d \end{bmatrix}$$

## 4. THE CALCULATION OF CORRELATIONCONSTRAINTS

*4.1 Auto-Correlation Constraint:*

In the conventional method for calculation of maximum non- zero shift auto-correlation as given in definition 2.1, the weighted bits' positions of code X are compared with circular shifted versions of code X. There are $n(n-1)$ comparisons of binary digits in the calculation of maximum non-zeroshift auto-correlation of uni-polar code as given in definition 2.1. The 'n' bits of code X are compared with 'n' bits of each of(n-1) circular shifted versions of code X. Thesecomparisons of weighted bits positions can be further reduced asdescribed below.

*Lemma 4.2:*

Incalculationofthemaximumnon-zeroshiftauto-correlation using weighted positions representation (WPR) of the code, thereare $(n-1)w$ comparisons of weighted positions (definition 2.3).

Proof: In conventional method for the calculation of the maximum non-zero shift auto-correlation of the uni-polar code, each of '$w$' weighted positions of $X_P$ are compared with each of '$w$' weighted positions of every (n-1) circular shifted versions ($X_P$+a), (Definition 2.3). Thus there are $(n-1)w^2$ comparisons of weighted positions in the calculation of the maximum non-zero shift auto-correlationof the code X.

*Lemma 4.3:*

For the uni-polar code of length '$n$' and weight '$w$', the total cases of overlapping pairs of weighted bits of uni-polar code with its circular shifted versions in the calculation of maximum non-zero shift auto-correlation (definition 2.1)are

$$\frac{w(w-1)}{2}.$$

Proof: In the calculation of maximum non-zero shiftauto-correlation, first weighted bit of the uni-polar code overlap with next ($w$-1) other weighted bits by circular shifting. The second weighted bit overlap with next ($w$-2) weighted bits by circular shifting. Similarly the third and so on up to ($w$-1)th weighted bit overlap with next ($w$-3) and so on up to last weighted bit by circular shifting. There are total ($w$-1) plus($w$-2) plus ($w$-3) plus and so on up to plus one overlapping which may occur in the pairs of codes with its maximum ($n$-1) circular shifted versions. Total overlapping of weighted bits are $w(w-1)/2$.

Lemma 4.4:

Theuni-polarcodewithcodelength'n'andcodeweight'w' has 'w' circular shifted versions with first bit as weighted bit of thecode.

*Example 4.1.4:* Let us take the code X =[0 1 0 1 0 0 1 0 0 0 1 0 0] with weighted positions representation $X_P$ = (1,3,6,10). The w=4 circular shifted versions of the code with first bit as weighted bit are given as follows

$X_1$= [1 0 1 0 0 1 0 0 0 1 0 0 0] , ($X_p$+12) = (0,2,5,9),

$X_3$= [1 0 0 1 0 0 0 1 0 0 0 1 0] , ($X_P$+10) = (0,3,7,11),

$X_6$= [1 0 0 0 1 0 0 0 1 0 1 0 0] , ($X_p$+7) = (0,4,8,10),

$X_{10}$= [1000101001000],($X_P$+3)=(0,4,6,9).Inthe calculation of maximum non-zero-shift auto-correlation, total cases of overlapping of weighted bits = first weighted bit overlaps with second, third and fourth weighted bits, i.e three overlapping + second weighted bit overlaps with third, and fourth, i.e. two overlapping + third weighted bit overlapswith fourth weighted bit, i.e. one overlapping = 6 overlapping.

*Lemma 4.5:*

There are${}^{w}C_2 = \frac{w(w-1)}{2}$ pairs of codes formed out of the 'w' circular shifted versions of the code with first bit as weighted bit.

*Lemma 4.6:*

Total unrepeated overlapping of weighted bits in the all pairs of codes having first bit as weighted bit are$\frac{w(w-1)}{2}$.

Proof: As per lemma 4.5, there are $\frac{w(w-1)}{2}$ pairs of codes havingfirstbitasweightedbit.Eachpairhasoneunrepeated overlapping at first position. Then, there are total $\frac{w(w-1)}{2}$ overlapping of weighted bits.

*Theorem 4.7:*

The overlapping of weighted bits of uni-polar code with its every circular shifted version equal to overlapping of weighted bits in all the pairs of the '*w*' circular shifted versions with first bit as weighted bit of thecode.

Proof: As per lemma 4.3, total overlapping of weighted bits of uni-polar code with its every circular shifted versions are $\frac{w(w-1)}{2}$. Asperlemma4.6,thesamenumberofdefinite overlapping are found in all the pairs of the '*w*' circular shifted versions with first bit as weighted bit of the code. Hence all weighted overlapping are covered in both cases.

*Theorem 4.8:*

The weighted positions of the 'w' circular shifted versions of the code with first bit as weighted bit are given by the rows of EdoP matrix along-with extra first column having zero elements.

Proof: Let us take the code X with DoPR($a$,$b$, $c$, $d$, $e$) with weight$w$=5,andcodelength$n$=($a$+$b$+$c$+$d$+$e$).The weighted positions of the code with first bit as weighted bitare (0, $a$, $a+b$, $a+b+c$, $a+b+c+d$). The circular shifted versions of this code with first bit as weighted bit are

(0,$b$,$b+c$,$b+c+d$,$b+c+d+e$)

(0, $c$, $c+d$, $c+d+e$, $c+d+e+a$) ,

( (0, $d$, $d+a$, $d+a+b$, $d+a+b+c$) and

(0, $e$, $e+a$, $e+a+b$, $e+a+b+c$) .

These circular shifted versions of the code with first bit as weighted bit are same as row element of the following EDoP matrix (example 3.7(a)) along-with extra first column having zero elements.

$$\begin{bmatrix} 0 & a & a+b & a+b+c & a+b+c+d \\ 0 & b & b+c & b+c+d & b+c+d+e \\ 0 & c & c+d & c+d+e & c+d+e+a \\ 0 & d & d+e & d+e+a & d+e+a+b \\ 0 & e & e+a & e+a+b & e+a+b+c \end{bmatrix}$$

Similarly for any weight $w \geq 2$ the theorem can be verified easily.

*Theorem 4.9:*

The maximum non-zero shift auto-correlation of the uni-polar code is equal to maximum number of overlapping bits among the pairs of 'w' circular shifted versions with first bit as weighted bit of the code. OR The maximum non-zero shift auto-correlation of the uni-polar code is equal to the maximum number of common DoP elements between two rows of EDoPmatrix having zero elements in first column.

$$\lambda_{ax} \geq \sum_{j=0}^{w-1} \sum_{l=0}^{w-1} e_{xij} e_{xkl} \; for i = (0{:}w-1), k = (i+1{:}w-1)$$

OR

The maximum non-zero shift auto-correlation of the uni-polar code is equal to one plus maximum number of common DoP elements between two rows of EDoP matrix of the code.

$$\lambda_{ax} \geq 1 + \sum_{j=0}^{w-1} \sum_{l=1}^{w-1} e_{xij} e_{ykl} \; for i = (0{:}w-1), l = (i+1{:}w-1)$$

where $e_{xij} e_{xkl} = \begin{cases} 1 if \; e_{xij} = e_{xkl} \\ 0 if \; e_{xij} = e_{xkl} \end{cases}$

$e_{xij}$ & $e_{xkl}$ are DoP elements of two rows of EDoP matrix along-with first column having zero elements.

Proof: Let us take the code X with DoPR($d_{x1}$, $d_{x2}$ ,..., $d_{xw}$) with weight '$w$' and code length $n = d_{x1} + d_{x2} + \ldots + d_{xw}$, the EDoP matrix with zero elements in the first column is formed asfollows.

$$\text{EDoP}\begin{bmatrix} e_{x00} & e_{x01} & e_{x02} & \cdots & e_{x0(w-1)} \\ e_{x10} & e_{x11} & e_{x12} & \cdots & e_{x1(w-1)} \\ e_{x20} & e_{x21} & e_{x22} & \cdots & e_{x2(w-1)} \\ \cdots & \cdots & \cdots & \cdots & \cdots \\ e_{x(w-1)0} & e_{x(w-1)1} & e_{x(w-1)2} & \cdots & e_{x(w-1)(w-1)} \end{bmatrix}$$

with

$e_{x00} = e_{x10} = e_{x20} = \ldots = e_{x(w-1)0} = 0.$

$e_{x01} = d_{x1}; e_{x11} = d_{x2}; \ldots; e_{x(w-1)1} = d_{xw}.$

$e_{x02} = d_{x1} + d_{x2}; e_{x12} = d_{x2} + d_{x3}; \ldots; e_{x(w-1)2} = d_{xw} + d_{x1}$

...

$e_{x0(w-1)} = d_{x1} + d_{x2} + \ldots + d_{xw-1}; e_{x1(w-1)} = d_{x2} + d_{x3} + \ldots + d_{xw}; \ldots;$

$e_{x(w-1)(w-1)} = d_{xw} + d_{x1} + d_{x2} + \ldots + d_{x(w-2)}.$

As per definition 2.1 and theorem 4.7, the maximum non-zero shift auto-correlation $\lambda_{ax}$ of the uni-polar code is equal to maximum number of overlapping of weighted bits among the pairs of circular shifted versions with first bit as weighted bit of the code. However As per theorem 4.8 and definition 2.3, the maximum non-zero shift auto-correlation of the code is equal to the maximum number of common DoP elements between two rows of EdoP matrix along-with firstcolumn with zero elements if $e_{xij}e_{xkl} = \begin{cases} 1 \; if e_{xij} = e_{xkl} \\ 0 \; if e_{xij} \neq e_{xkl} \end{cases}$ which represents common elements between two rows of EdoP matrix with first column having zero elements.

$$\lambda_{ax} \geq \sum_{j=0}^{w-1} \sum_{l=0}^{w-1} e_{xij}e_{xkl} \; for i = (0{:}\, w-1), l = (i+1{:}\, w-1)$$

Or the maximum non-zero shift auto-correlation of the uni-polar code is equal to one plus maximum common DoP elements between two rows of EDoP matrix of the code. As any two rows of EdoP matrix with first column having zero elements always has at least one common element which is zero.

$$\lambda_{ax} \geq 1 + \sum_{j=0}^{w-1} \sum_{l=1}^{w-1} e_{xij}e_{ykl} \; for i = (0{:}\, w-1), l = (i+1{:}\, w-1)$$

*Lemma4.10:*

**T**here willbe $\frac{(w-1)w^3}{2}$comparisons of DoP elements in the calculation of maximum non-zero shift auto-correlation using extended DoP matrix with first column having zero elements.

Proof: There are $\frac{(w-1)w}{2}$pair of rows of extended DoP matrix which are compared in the calculation of maximum non-zero shift auto-correlation of the code. In each pair of rows, there are $w^2$ comparisons of DoP elements. Thus there are total $\frac{(w-1)w^3}{2}$comparisons of DoP elements of EDoP matrix take place in the calculation of maximum non-zero shift auto-correlation of the code.

*Lemma 4.11:*

There are $\frac{w(w-1)^3}{2}$ comparisons of DoP elements in the calculation of maximum non-zero shift auto-correlation using extended DoP matrix of the code.

Proof: There are $\frac{(w-1)w}{2}$ pair of rows of extended DoP matrix which are compared in the calculation of maximum non-zero shift auto-correlation of the code. In each pair of rows, there are $(w-1)^2$ comparisons of DoP elements. Hence there are $\frac{w(w-1)^3}{2}$ total comparisons of DoP elements of EDoP matrix in the calculation of maximum non-zero shift auto-correlation of the code.

*Lemma 4.12:*

If there is no common DoP elements in the pair of rows of EDoP matrix of code, the maximum non-zero shift auto-correlation of the code is always equals to one [38].

*4.13 Cross-Correlation Constraint:*

In the conventional method for calculation of cross-correlation for the pair of uni-polar codes as given in definition 2.6, the weighted bits' positions of code X are compared with code Y and circular shifted versions of code Y. Or the weighted bits' positions of code Y are compared with code X and circular shifted versions of code X. There are $n^2$ comparisons of binary digits in the calculation of cross-correlation of uni-polar code in conventional method (definition 2.6).

These comparisons of weighted bits positions can be further reduced as described below.

*Lemma 4.14:*

In the calculation of cross-correlation using weighted positions representation (WPR) of the pair of codes, there are ($nw^2$ ) comparisons of weighted positions (definition 2.8).

Proof: In the calculation of cross-correlation of the pair of uni- polar codes (definition 2.8), the '$w$' weighted positions (WP) of $X_P$ are compared with '$w$' weighted positions of $Y_P$ and each of the (n-1) circular shifted versions ($Y_P$+a).There are($nw^2$ ) total comparisons of weighted position in the calculation of cross-correlation of the pair of codes.

*Lemma 4.15:*

For the uni-polar codes of length '$n$' and weight '$w$', the definite cases of overlapping of weighted bits of uni-polar code X with code Y and the ($n$-1) circular shifted versions of code Y are $w^2$ .

Proof: In the calculation of cross-correlation (definition 2.6), first weighted bit of code X overlap with $w$ weighted bits of code Y in '$w$' shifts. The second weighted bit of code X overlap with '$w$' weighted bit of code Y in '$w$' shifts. Similarly the third and so on upto $w^{th}$ weighted bit of code X overlap with '$w$' weighted positions of code Y in '$w$' shifts. Thus there are ($w^2$) total overlapping of weighted bits occurred in the pairs of code X with code Y and the maximum ($n$-1) circular shifted versions of code Y in the calculation of cross-correlation.

*Lemma 4.16:*

There are total $w^2$ pairs of code X and code Y formed out of the 'w' circular shifted versions of both the codes with first bit as weighted bit.

*Lemma 4.17:*

The definite overlapping of weighted bits in all the pairs of circular shifted versions of codes X and Y having first bit as weighted bit are $w^2$ .

Proof: As per lemma 4.16, there are $w^2$ pairs of codes having first bit as weighted bit. Each pair has one definite overlapping at first position. Subsequently there are $w^2$ definite overlapping of weighted bits.

Theorem4.18: The overlapping of weighted bits of uni-polar code X with uni-polar code Y and every circular shifted version of code Y equals to the overlapping of weighted bits in all the pairs of code X and code Y formed out of the '$w$' circular shifted versions of both the codes having first bit as weighted bit.

Proof: As per lemma 4.15, the definite overlapping of weighted bits of uni-polar code X and code Y along-with every circular shifted version of code Y are $w^2$ . As well as per lemma 4.17, the same number of definite overlapping are covered in all the pairs of circular shifted versions of code X and code Y having first bit as weighted bit. Hence all definite weighted overlapping are covered in both cases.

*Theorem 4.19:*

The cross-correlation of the uni-polar codes X and Y is equal to maximum overlapping among the pairs of code X and code Y out of the '$w$' circular shifted versions with first bit as weighted bit of both the codes.

OR

The cross-correlation of the uni-polar codes X and Y is equal to maximum common DoP elements between any two rows of EdoP matrices along-with first column with zero elements of code X and code Y respectively.

$$\lambda_{cxy} \geq \sum_{j=0}^{w-1} \sum_{l=0}^{w-1} e_{xij} e_{ykl} \, for i = (0{:}\, w-1), k = (0{:}\, w-1)$$

OR

The cross-correlation of the uni-polar codes X and Y is equal to one plus maximum common DoP elements between any two rows of EDoP matrices of code X and code Y respectively.

$$\lambda_{cxy} \geq 1 + \sum_{j=1}^{w-1} \sum_{l=1}^{w-1} e_{xij} e_{ykl} \, for i = (0{:}\, w-1), k = (0{:}\, w-1)$$

if $e_{xij} e_{ykl} = \begin{cases} 1 \; if e_{xij} = e_{ykl} \\ 0 \; if e_{xij} \neq e_{ykl} \end{cases}$

$e_{ij}$&$e_{kl}$are DoP elements of the rows of EDoP matrices along-with extra column with zero elements of code X and code Y respectively.

Proof: Suppose the code X with DoPR$(d_{x1}, d_{x2} \dots \dots, d_{xw})$and code Y with DoPR$\left(d_{y1}, d_{y2} \dots \dots, d_{yw}\right)$with weight '$w$' and code length $n = d_{x1} + d_{x2} + \cdots \dots + d_{xw} = d_{y1} + d_{y2} + \cdots \dots + d_{yw}$the EDoP matrix along-with first column with zero elements of code X and code Y are formed as follows

$$\text{EDoP(X)}\begin{bmatrix} e_{x00} & e_{x01} & e_{x02} & \cdots & e_{x0(w-1)} \\ e_{x10} & e_{x11} & e_{x12} & \cdots & e_{x1(w-1)} \\ e_{x20} & e_{x21} & e_{x22} & \cdots & e_{x2(w-1)} \\ \cdots & \cdots & \cdots & \cdots & \cdots \\ e_{x(w-1)0} & e_{x(w-1)1} & e_{x(w-1)2} & \cdots & e_{x(w-1)(w-1)} \end{bmatrix}$$

with

$e_{x00} = e_{x10} = e_{x20} = \ldots = e_{x(w-1)0} = 0.$

$e_{x01} = d_{x1}; e_{x11} = d_{x2}; \ldots; e_{x(w-1)1} = d_{xw}.$

$e_{x02} = d_{x1} + d_{x2}; e_{x12} = d_{x2} + d_{x3}; \ldots; e_{x(w-1)2} = d_{xw} + d_{x1}$

...

$e_{x0(w-1)} = d_{x1} + d_{x2} + \ldots + d_{x(w-1)}; e_{x1(w-1)} = d_{x2} + d_{x3} + \ldots + d_{xw}; \ldots;$

$e_{x(w-1)(w-1)} = d_{xw} + d_{x1} + d_{x2} + \ldots + d_{x(w-2)}.$

and

$$\text{EDoP(Y)}\begin{bmatrix} e_{y00} & e_{y01} & e_{y02} & \cdots & e_{y0(w-1)} \\ e_{y10} & e_{y11} & e_{y12} & \cdots & e_{y1(w-1)} \\ e_{y20} & e_{y21} & e_{y22} & \cdots & e_{y2(w-1)} \\ \cdots & \cdots & \cdots & \cdots & \cdots \\ e_{y(w-1)0} & e_{y(w-1)1} & e_{y(w-1)2} & \cdots & e_{y(w-1)(w-1)} \end{bmatrix}$$

*OR*

$e_{y00} = e_{y10} = e_{y20} = \ldots = e_{y(w-1)0} = 0.$

$e_{y01} = d_{y1}; e_{y11} = d_{y2}; \ldots; e_{y(w-1)1} = d_{yw}.$

$e_{y02} = d_{y1} + d_{y2}; e_{y12} = d_{y2} + d_{y3}; \ldots; e_{y(w-1)2} = d_{yw} + d_{y1}$

...

$e_{y0(w-1)} = d_{y1} + d_{y2} + \ldots + d_{y(w-1)}; e_{y1(w-1)} = d_{y2} + d_{y3} + \ldots + d_{yw}; \ldots;$

$e_{y(w-1)(w-1)} = d_{yw} + d_{y1} + d_{y2} + \ldots + d_{y(w-2)}.$

As per definition 2.6 and theorem 4.18, the cross-correlation$\lambda_{cxy}$ of the uni-polar codes X and Y is equal to the maximum number of overlapping among the pairs of code X and code Y of the circular shifted versions with first bit as weighted bit of both the codes. However As per definition 2.6 and theorem 4.8, the cross-correlation of the codes X and Y is the maximum number of common DoP elements between the rows of EDoP matrices along-with first column having zero elements for both the codes X and Y.

$$\text{if } e_{xij} e_{ykl} = \begin{cases} 1 \; if e_{xij} = e_{ykl} \\ 0 \; if e_{xij} \neq e_{ykl} \end{cases}$$

$$\lambda_{cxy} \geq \sum_{j=0}^{w-1} \sum_{l=0}^{w-1} e_{xij} e_{ykl} \; fori = (0{:}w-1), k = (0{:}w-1)$$

$e_{ij}$ &$e_{kl}$are DoP elementsare the DoP elements of the rows of EDoP matrices along-with

first extra column having zero elements of code X and code Y respectively..

Or the cross-correlation of the uni-polar codes X and Y is equal to one plus maximum common DoP elements between the rows of EDoP matrices of the code X and Y. Because any two rows of EdoP matrices along-with first column having zero elements for code X and Y always has at least one common element as zero.

$$\lambda_{cxy} \geq 1 + \sum_{j=1}^{w-1} \sum_{l=1}^{w-1} e_{xij} e_{ykl} \; for i = (0{:}\, w-1), k = (0{:}\, w-1)$$

*Lemma 4.20:*

There are $w^4$ comparisons of DoP elements in the calculation of cross-correlation using extended DoP matrices along-with first column having zero elements for both the codes,.

Proof: As per lemma 4.16, there are $w^2$ pair of rows from extended DoP matrices along-with first column having zero elements for code X and Y. In each pair of rows, there are $w^2$ comparisons of DoP elements. Hence there are $w^4$ comparisons of DoP elements of both EDoP matrices in the calculation of cross-correlation for the pair of codes X and Y.

*Lemma 4.21:*

In calculation of cross-correlation using extended DoPmatrices of the codes X and Y, there are $w^2 \, (w-1)^2$ comparisons of DoP elements.

Proof: As per lemma 4.16, there are $w^2$ pair of rows fromextended DoP matrices of code X and Y. In each pair ofrows,there are $(w-1)^2$ comparisons of DoP elements. Hencethere $w^2(w-1)^2$ total comparisons of DoP elements of EDoP matrices in the calculation of cross-correlation of the codes X with Y.

*Lemma 4.22:*

If there is no common DoP elements in the pair of rows of EDoP matrices of the two codes, the cross-correlation of the pair of codes is always equals to one [38].

*Theorem 4.23:*

The cross-correlation of the uni-polar code X with code parameters ($n_1$, $w_1$, $\lambda_{a1}$ ) and code Y with parameters($n_2$ , $w_2$ , $\lambda_{a2}$ ) is equal to maximum common DoP elementsbetween the any two rows of EdoP matrices along-with first column having zero elements of code X and code Yrespectively.

$$\lambda_{cxy} \geq 1 + \sum_{j=1}^{w-1} \sum_{l=1}^{w-1} e_{xij} e_{ykl} \; for i = (0{:}\, w_1 - 1), k = (0{:}\, w_2 - 1)$$

$$\text{if } e_{xij} e_{ykl} = \begin{cases} 1 \; if e_{xij} = e_{ykl} \\ 0 \; if e_{xij} \neq e_{ykl} \end{cases}$$

$e_{ij}$ &$e_{kl}$areDoP elements of the rows of EDoP matrices along-with first column having zero elements of code X and code Y respectively.

Proof: it is straight forward through theorem 4.19.

## I. DESIGN OF THE MAXIMAL SETS OF1-DUOC

The maximal sets of 1-DUOC for fixed code parameters ($n$, $w$, $\lambda_a$ =1, $\lambda_c$=1) are designed through maximal clique search method in the proposedalgorithm.

*5.1. Maximal Clique Search Method:*

Suppose the uni-polar code in standard DoPRis($d_1$,$d_2$,...,$d_w$). Let for $u<w$, ($d_1$,$d_2$,...,$d_u$) DoPelements of the code are known at one step of following algorithm. The extended DoP matrix of the code withincomplete DoP elements ($d_1$, $d_2$ ,..., $d_u$ ) is given by lemma

3.10. The maximum non-zero shift auto-correlation and cross- correlation of the pair of codes with incomplete DoP elements can be calculated by theorem 4.9 and theorem 4.19 respectively as well as using lemma 3.10 for EDoP matrices of codes with incomplete DoP elements. In the defined graph of codes with incomplete or complete DoP elements, some maximal clique sets of codes are searched by maximal clique finding algorithms [41-46] or the algorithm proposed here as follows.

*Step.1:*

Define the connected graph of codes or correlation matrix having binary elements (0,1). In the correlation matrix of codes, the binary digit '1' represent to cross-correlation lessthan equal to $\lambda_c$between two codes while binary digit '0'represent to cross-correlation greater than $\lambda_c$ between two codes in the matrix. In the defined graph of codes, the binary digit '1' of correlation matrix is equivalent to a straight line between two codes while no line if corresponding binary digit '0' for two codes in correlation matrix. i=0. If the size of correlation matrix is large enough, the correlation matrix can be broken into smaller parts which is discussed in the next section. All the codes of graph can be divided into S sub-graphs. Each sub-graph is containing the codes with similar DoP elements of some positions which is clearly mentioned in next sub-section 5.2.

*Step.2:*

In the defined graph or smaller part of correlation matrix with binary elements, find the codes with highest degree. The smaller part of correlation matrix contains the codes of $(i+1)^{th}$ sub-graph along the rows of matrix while the codes of $(i+2)^{th}$ and other subsequent sub-graphs found along the columns, select one of the codes with highest degree. Put this code inset

A. i=i+1. d=highest degree.

*Step.3:*

If d>1, form another graph or smaller part of correlation matrix containing only the codes connected with highest degree code selected in step-2 at last time and excluding the codes of set A. Jump to step-2

*Step.4:*

If d=1, find the code adjacent with last selected highest degree code as in step-2. Put this code also in set A. The codes of set A form a maximal clique defined by Lemma 2.3. There are total 'i' codes are found in final set A.

*Step.5:*

More than one maximal clique sets can be searched out of first smaller part of correlation matrix having more than one codes with highest degree by following the step 1 to step 4. The first smaller part of correlation matrix contains the codes of $1^{st}$ sub- graph along the rows while the codes of $2^{nd}$ and subsequent sub-graphs found along the columns.

*Step.6:*

Out of the all the M maximal clique sets found at step 5, a clique sets correlation matrix MxM can be defined. This matrix contains the elements equivalentto cross-correlation of pair of maximal clique sets. The cross-correlation of a pair of maximal clique sets is defined as maximumcross- correlation between a pair of codes taken from each maximal clique set of pair. The clique sets correlation matrix can be normalized with elements equal to 'zero' if cross-correlation value is greater than ($\lambda_c$ +1) and normalized elements equal to 'one' if cross-correlation value is less than equal to ($\lambda_c$ +1).

*Step.7:*

By applying step 2 to step 4 over normalized clique set correlation matrix, not for smaller part of correlation matrix as given for the codes, one set of maximal clique sets is searched out having minimum correlated maximal clique sets.

*5.2. Algorithm to design minimum correlated maximal cliquesest of 1-DUOC:*

*Step.1:*

Inputcodeparameters($n$=$n_1$,$w$=$w_1$,$\lambda_a$=1,$\lambda_c$=1)Such that$n$>>$w$>($\lambda_a$,$\lambda_c$).

*Step.2:*

For the code in standard DoPR ($d_1$, $d_2$ ,..., $d_{(w-1)}$ , $d_w$ ) , all possible pairs of unequal values of ($d_1$, $d_2$ ) from the range of ($d_1$, $d_2$ ) given in lemma 3.5 are arranged with serial number of codes with first only two DoP elements. Themaximum non-zero shift auto-correlation of the code withincomplete DoP elements should not greater than one. Any two such codes are defined as related if at least one common DoP elements found otherwise unrelated. A connected graph G can be defined having vertices equal to number of codes formed. One pair of ($d_1$, $d_2$ ) is equivalent to one node or code as well as unrelated codes are equivalent to line drawn between the nodes or code numbers. The codes are related if mutual cross- correlation is greater than one and unrelated if cross- correlation is equal toone.

The counter s=0.

*Step.3:*

From the defined graph G, a set of minimum correlated maximal clique sets is searched by maximal clique search method given at section 5.1. The counter s=s+1. If ($s$+2)=($w$−1)gotostep-6.

*Step.4:*

All the codes found with each maximal clique set found at step 3, are rearranged along-with (s+2)$^{th}$DoP element which is not equal to last (s+1)$^{th}$DoP elements of the code. Each graph of codes corresponding to maximal clique set is redefined as in step-2 such that only those codes are included in the graph having maximum non-zero shift auto-correlation equal to one. The maximum non-zero shift auto-correlation of the code, equal to one means no DoP elements in the rows of EDoP matrix of the code, are repeated in the same EDoP matrix of the code.

*Step.5:*

If $(s + 2) < (w - 1)$ , go to step-3 and continue.

*Step.6:*

Each code of last searched each maximal clique set is rearranged with its last DoP element as per theorem-3.3. i.e. $d_w$=$n$−($d_1$+$d_2$+...+$d_{(w-1)}$).

This searched maximal set is group of minimum correlated ($\lambda_c$ = 2) maximal clique sets of 1-DUOC with ($n$=$n_1$,$w$=$w_1$,$\lambda_a$=1,$\lambda_c$=1).

*Step.7:*

Similarly for ($n$ =($n_1$, $n_2$ , $n_3$ ,...), $w$ =($w_1$, $w_2$ , $w_3$ ,...), $\lambda_a$ =1, $\lambda_c$=1) minimum correlated multiple maximal sets of codes are formedforeachsetofcodeparameters($n$,$w$,$\lambda_a$,$\lambda_c$)asin step-1 tostep-6.

*Step.8:*

Now another graph F is defined for sets of 1-DUOC with different code parameters. The vertices of graph are equivalent to sets of 1-DUOC with different set of code parameters. The line drawn between two vertices is equivalent to the cross- correlation between two sets which is less than or equal to two. No line drawn between two vertices if the two equivalent sets have cross-correlation greater than two. In this defined graph F of sets of 1-DUOC with different set of code parameters, a maximal clique is searched out by algorithm given in sub-section 5.1. This searched clique contains the sets of 1-DUOC with multi-length and multi-weight having auto- correlation constraint equal to one and cross-correlation constraintequaltotwo.Thealgorithmproposedinsubsection 5.2 can design sets of codes with specified auto (cross) correlation constraint after some little change in thealgorithm.

## 5. CONCLUSION

In this paper the proposed clique search algorithm design the family of sets of codes with multi-length, multi-weight, auto-correlation constraint equal to one and cross-correlation constraint equal to one for within set while cross-correlation constraint equal to two among the sets with upper bound. These codes are designed for unspecific code parameters which increase inherent security. These codes can be utilized for the purpose of increasing the channel capacity and even for multi-rate system incorporating OOC. The computational complexity of the proposed algorithm designing the multiple sets of codes with variable and general code parameters, is polynomial type if clique search is polynomial. In future a mathematical alternative of clique search method for 1-DUOC or OOC may be explored for reducing the computational complexity of the algorithm proposedhere.